\documentclass[aps,prb,twocolumn,floatfix,superscriptaddress]{revtex4-1}
\usepackage{graphicx,txfonts,bm}
\usepackage[colorlinks,citecolor=magenta,linkcolor=blue]{hyperref}

\begin{document}


\title{Temperature-driven hidden 5$f$ itinerant-localized crossover in heavy-fermion compound PuIn$_3$}


\author{Haiyan Lu}
\email{hyluphys@163.com}
\affiliation{Science and Technology on Surface Physics and Chemistry Laboratory, P.O. Box 9-35, Jiangyou 621908, China}

\author{Li Huang}
\affiliation{Science and Technology on Surface Physics and Chemistry Laboratory, P.O. Box 9-35, Jiangyou 621908, China}

\date{\today}


\begin{abstract}
The temperature-dependent evolution pattern of 5$f$ electrons helps to elucidate the long-standing itinerant-localized dual nature in plutonium-based compounds. 
In this work, we investigate the correlated electronic states of PuIn$_3$ dependence on temperature by using a combination of the density functional theory and the dynamical mean-field theory. Not only the experimental photoemission spectroscopy is correctly reproduced, but also a possible hidden 5$f$ itinerant-localized crossover is identified.
Moreover, it is found that the quasiparticle multiplets from the many-body transitions gradually enhance with decreasing temperature, accompanied by the hybridizations with 5$f$ electrons and conduction bands. 
The temperature-induced variation of Fermi surface topology suggests a possible electronic Lifshitz transition and the onset of magnetic order at low temperature. Finally, the ubiquitous existence orbital selective 5$f$ electron correlation is also discovered in PuIn$_3$. These illuminating results shall enrich the understanding on Pu-based compounds and serve as critical predictions for ongoing experimental research. 
\end{abstract}


\maketitle

\section{Introduction\label{sec:intro}}
Plutonium (Pu) situates on the edge between the obviously hybridized 5$f$ states of uranium~\cite{PhysRevLett.44.1612} and mostly localized 5$f$ states of americium~\cite{smith5636}, signifying the itinerant-localized nature of 5$f$ electrons~\cite{LAReview,RevModPhys.81.235}. The spatial extension of partially filled 5$f$ states enables the hybridization with conduction bands, facilitating active chemical bonding and formation of abundant plutonium-based compounds. Besides the fantastic physical properties of Pu which are governed by the 5$f$ states~\cite{albers:2001,Hecker2004,PhysRevB.72.054416,Janoscheke:2015,PhysRevX.5.011008,dai:2003,wong:2003}, the Pu-based compounds demonstrate novel quantum phenomena including unconventional superconductivity~\cite{PhysRevLett.108.017001,PhysRevLett.91.176401,Sarrao2002,Curro2005,Daghero2012}, nontrivial topology~\cite{PhysRevLett.111.176404,PhysRevB.99.035104}, complicated magnetic order~\cite{Chudo2013}, and heavy-fermion behavior~\cite{Bauer2015Plutonium}, to name a few. 
The discovery of superconductivity in PuCoGa$_5$~\cite{PhysRevLett.91.176401,Sarrao2002,Curro2005,Daghero2012} with astonishingly high transition temperature of 18.5 K has renewed an interest in Pu-based compounds. The unconventional superconductivity in Pu-based ``115'' system (Pu$M$$X_{5}$, $M$=Co, Rh; $X$=Ga, In) is intimately intertwined with 5$f$ electrons which manifest themselves in plenty of ground state properties comprising magnetism, superconductivity, and charge density wave~\cite{PhysRevLett.93.147005,PhysRevLett.94.016401}. 

\begin{figure}[ht]
\centering
\includegraphics[width=\columnwidth]{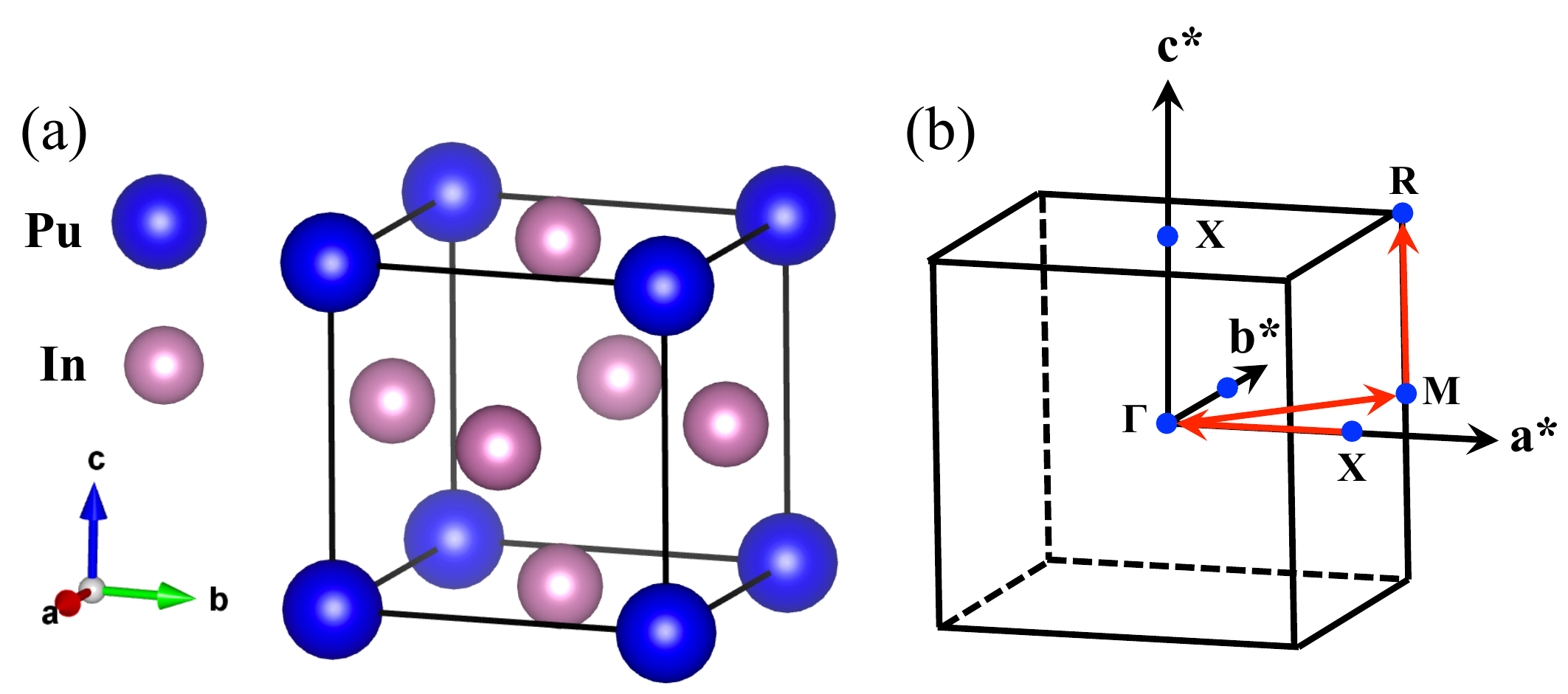}
\caption{(Color online). (a) Crystal structure of PuIn$_3$. (b) Schematic picture of the first Brillouin zone of PuIn$_3$. Some high-symmetry $k$ points $X$ [0.5, 0.0, 0.0], $\Gamma$ [0.0, 0.0, 0.0], $M$ [0.5, 0.5, 0.0], and $R$ [0.5, 0.5, 0.5] are marked.
\label{fig:tstruct}}
\end{figure}

PuIn$_3$ crystallizes in cubic AuCu$_{3}$ structure (space group $Pm$-3$m$) [see Fig.~\ref{fig:tstruct}(a)] with lattice constant 4.703 \AA~\cite{Eric2466}, which is the parent material of PuCoIn$_5$ with the insertion of the CoIn$_2$ layer into the cubic structure.
PuIn$_3$ is a paradigm Pu-based heavy-fermion compound~\cite{Bauer2015Plutonium} with an electronic specific heat coefficient of 307 mJ/(mol$\times$K$^2$), suggesting a substantial effective mass enhancement. Moreover, the measured temperature dependence of electrical resistivity decreases rapidly below 50 K, displaying representative heavy-fermion behavior. Since the Pu-Pu distance in PuIn$_3$ is larger than the Hill limit, the specific heat, electrical resistivity, magnetic susceptibility and $^{115}$In nuclear quadrupole resonance~\cite{Chudo2013} experiments all indicate the onset of antiferromagnetic order (AFM) below $T_N$=14.5 K. So far, the commensurate AFM and its underlying mechanism is controversial and deserves further clarification~\cite{Chudo2013,PhysRevB.88.125106}. 
Meanwhile, the de Haas-van Alphen (dHvA) oscillation identifies a Fermi surface pocket near the [111] direction with an enhanced cyclotron effective mass both in paramagnetic phase~\cite{JPSJ.74.2889,HAGA2007114} and antiferromagnetic state~\cite{2013Shubnikov}. The impact of magnetism on heavy-fermion state with effective mass renormalization is another interesting issue. It is notable that the heavy-fermion state and antiferromagnetic order are closely related to the intricate electronic structure of PuIn$_3$. The experimental photoemission spectroscopy of PuIn$_3$ above $T_N$ shows one peak around the Fermi level and the main peak at -1.2 eV, combined with mixed level model calculation within density functional theory (DFT), implying the itinerant-localized dual nature of 5$f$ electrons~\cite{JOYCE2006920}. On the theoretical side, the electronic structure, three-dimensional Fermi surface, dHvA quantum oscillation frequency and effective band mass in both the paramagnetic and antiferromagnetic states of PuIn$_3$ have been systematically addressed based on the density functional theory with a generalized gradient approximation~\cite{PhysRevB.88.125106}. Even though great efforts have been made to gain valuable insight into the itinerant-localized nature of 5$f$ states.
Only one of the three calculated Fermi surfaces on the basis of the 5$f$-itinerant band model~\cite{JPSJ.74.2889} has been observed by dHvA experiment. Besides, the experimental photoemission spectroscopy is well reproduced by employing mixed level model~\cite{JOYCE2006920} seems not quite consistent with the subsequent calculations~\cite{PhysRevB.88.125106}. The inconsistency might come from different approximations in traditional first-principles approaches. The itinerant degree of 5$f$ states is not quantitatively described without rigorous treating many-body effects. In addition, the ground state calculation fails to access the finite temperature effects. It is instructive to study the evolution of electronic structure as a function of temperature to probe the low-temperature magnetic state. Furthermore, large spin-orbital coupling and complicated magnetic ordering states make the theoretical calculations rather difficult. Accordingly, it seems tough to acquire an accurate and comprehensive picture for the electronic structure of PuIn$_3$.

In the paper, we present the electronic structure of PuIn$_3$ dependence on temperature using the density functional theory in combination with the single-site dynamical mean-field theory (DFT + DMFT). A comparative study of PuIn$_3$ shall shed light on the bonding behavior and correlation effects, so as to develop a deeper understanding of the relationships between electronic structure and antiferromagnetic state. We endeavor to elucidate the itinerant-localized 5$f$ states. We calculate the momentum-resolved spectral functions, density of states, Fermi surface, self-energy functions and valence state fluctuations of PuIn$_3$. It is discovered that 5$f$ states become itinerant at low temperature accompanied by moderate valence state fluctuations. Moreover, the change of Fermi surface topology possibly implies the development of antiferromagnetic order. Finally, it is found that strongly correlated 5$f$ electrons are orbital dependent, which seems commonly exists in Pu-based compounds.

The rest of this paper is organized as follows. In Sec.~\ref{sec:method}, the computational details are introduced. In Sec.~\ref{sec:results}, the electronic band structures, total and partial 5$f$ density of states, Fermi surface topology, 5$f$ self-energy functions, and probabilites of atomic eigenstates are presented. In Sec.~\ref{sec:dis}, we attempt to clarify the 4$f$ and 5$f$ electrons in isostructural compounds CeIn$_3$ and PuIn$_3$. Finally, Sec.~\ref{sec:summary} serves as a brief conclusion.


\section{Methods\label{sec:method}}
The well-established DFT + DMFT method combines realistic band structure calculation by DFT with the non-perturbative many-body treatment of local interaction effects in DMFT~\cite{RevModPhys.68.13,RevModPhys.78.865}. Here we perform charge fully self-consistent calculations to explore the detailed electronic structure of PuIn$_3$ using DFT + DMFT method. The implementation of this method is divided into DFT and DMFT parts, which are solved separately by using the \texttt{WIEN2K} code~\cite{wien2k} and the \texttt{EDMFTF} package~\cite{PhysRevB.81.195107}. 

In the DFT calculation, the experimental crystal structure of PuIn$_3$~\cite{Eric2466} was used. Since the calculated temperature is above the antiferromagnetic transition temperature, the system was assumed to be nonmagnetic. The generalized gradient approximation was adopted to formulate the exchange-correlation functional~\cite{PhysRevLett.77.3865}. The spin-orbit coupling was taken into account in a second-order variational manner. The $k$-points mesh was $15 \times 15 \times 15$ and $R_{\text{MT}}K_{\text{MAX}} = 8.0$.  

In the DMFT part, 5$f$ orbitals of plutonium were treated as correlated. The four-fermions interaction matrix was parameterized using the Coulomb interaction $U = 5.0$~eV and the Hund's exchange $J_H=0.6$~eV via the Slater integrals~\cite{PhysRevB.59.9903}. The fully localized limit scheme was used to calculate the double-counting term for impurity self-energy function~\cite{jpcm:1997}. The vertex-corrected one-crossing approximation (OCA) impurity solver~\cite{PhysRevB.64.115111} was employed to solve the resulting multi-orbital Anderson impurity models. 
 Note that we not only utilized the good quantum numbers $N$ (total occupancy) and $J$ (total angular momentum) to classify the atomic eigenstates, but also made a severe truncation ($N \in$ [3, 7]) for the local Hilbert space~\cite{PhysRevB.75.155113} to reduce the computational burden. The convergence criteria for charge and energy were $10^{-5}$ e and $10^{-5}$ Ry, respectively. It is worth mentioning that the advantage of OCA impurity solver lies in the real axis self-energy, the direct output $\Sigma (\omega)$ were applied to calculate the momentum-resolved spectral functions $A(\mathbf{k},\omega)$ and density of states $A(\omega)$, as well as other physical observables.

\section{Results of PuIn$_3$\label{sec:results}}

\begin{figure*}[th]
\centering
\includegraphics[width=\textwidth]{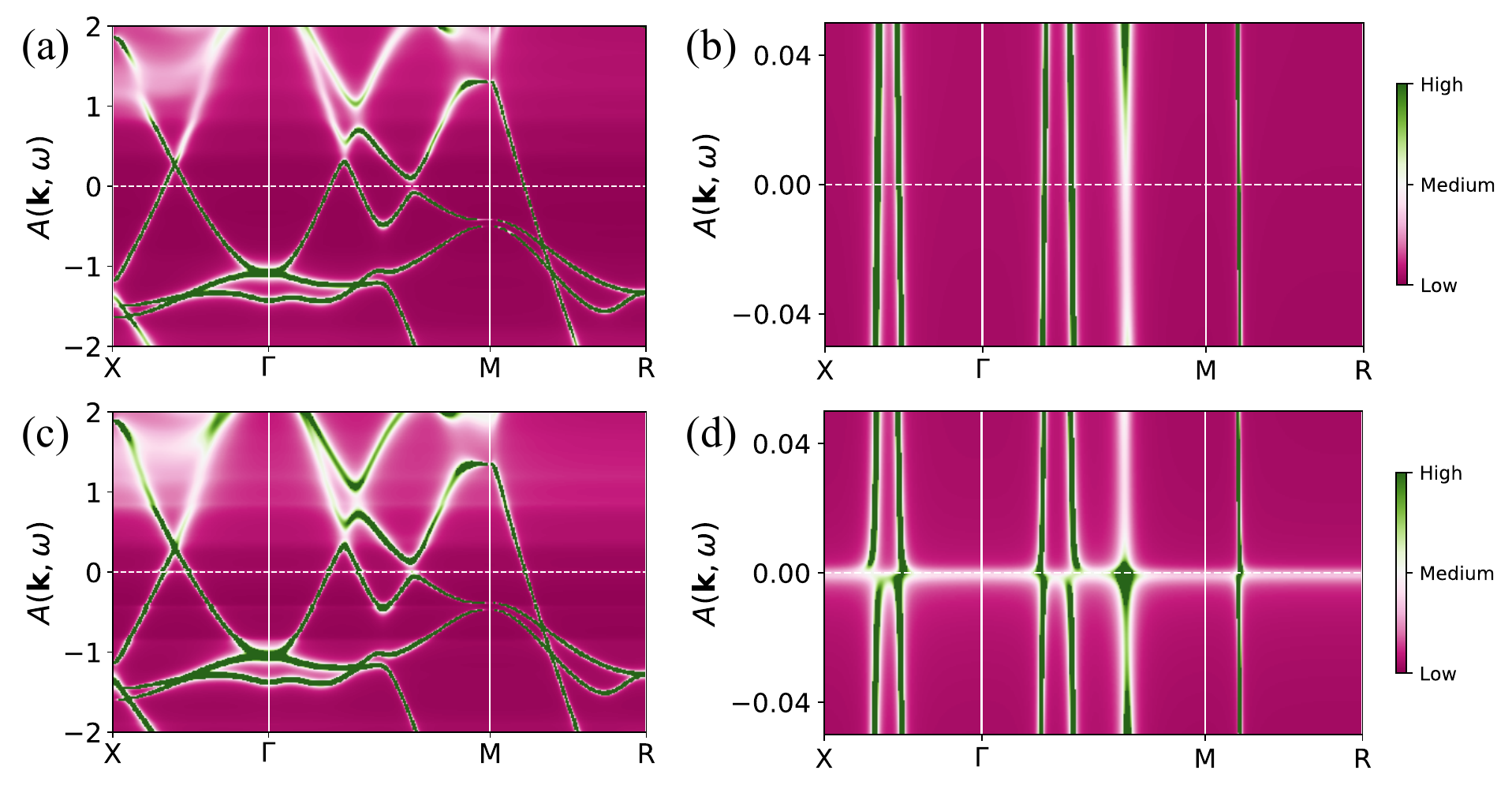}
\caption{(Color online). Momentum-resolved spectral functions $A(\mathbf{k},\omega)$ of PuIn$_3$ as a function of temperature under ambient pressure obtained by DFT + DMFT calculations. (a) 580 K. (c) 14.5 K. An enlarged view of panel (b) 580 K and (d) 14.5 K in the energy window $\omega \in [-0.05, 0.05]$ eV. In these panels, the horizontal lines denote the Fermi level.
\label{fig:akw}}
\end{figure*}

\begin{figure}[t]
\centering
\includegraphics[width=\columnwidth]{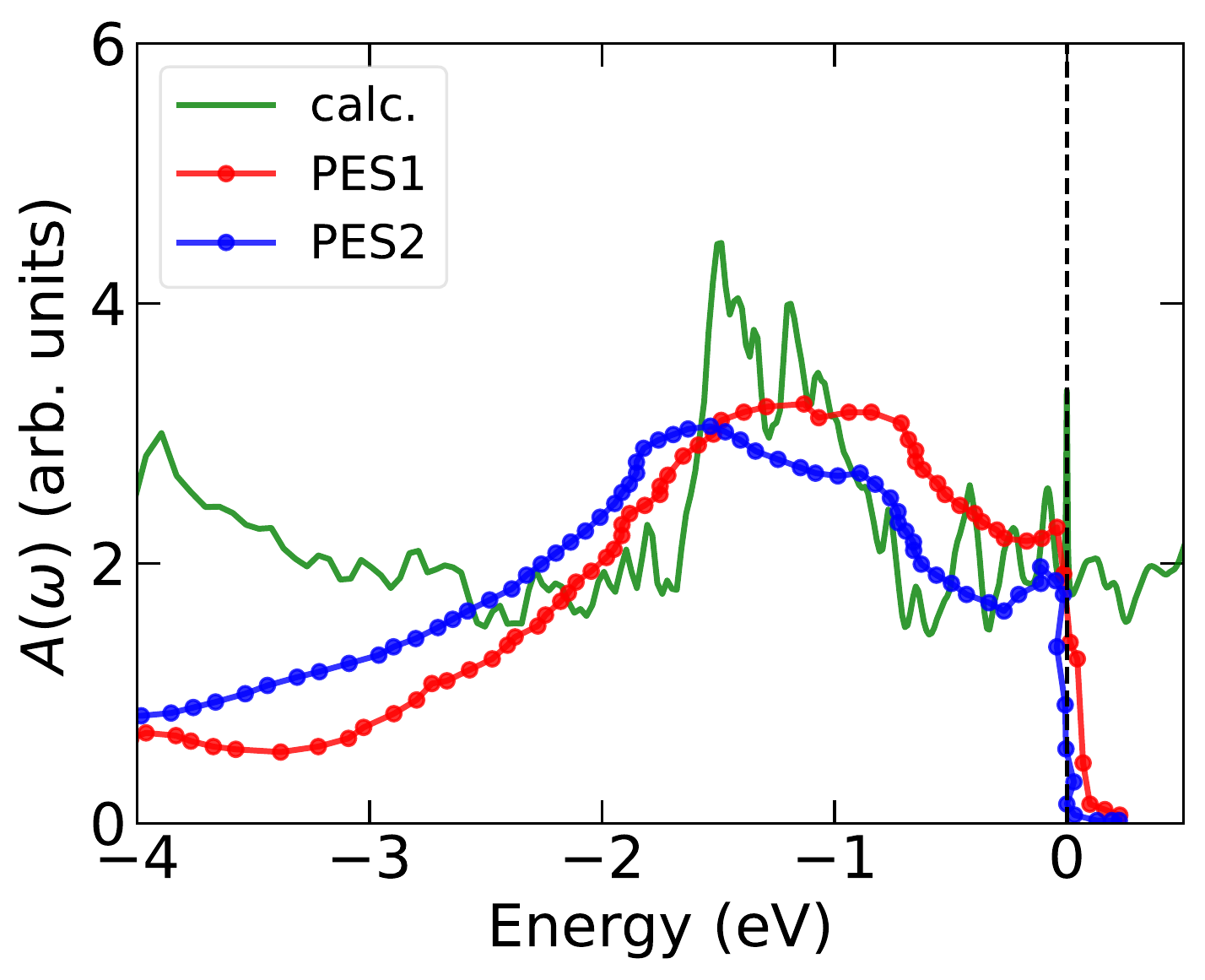}
\caption{(Color online). Electronic density of states of PuIn$_3$. The calculated and experimental data are represented by solid thick line and circles, respectively. The calculated data is multiplied by the Fermi-Dirac distribution function. The experimental data are extracted from Ref.~[\onlinecite{JOYCE2006920}] at two photon energies of 21.2 eV (red circles) and 40.8 eV (blue circles), respectively. \label{fig:tdos_exp}}
\end{figure}

\begin{figure*}[th]
\centering
\includegraphics[width=\textwidth]{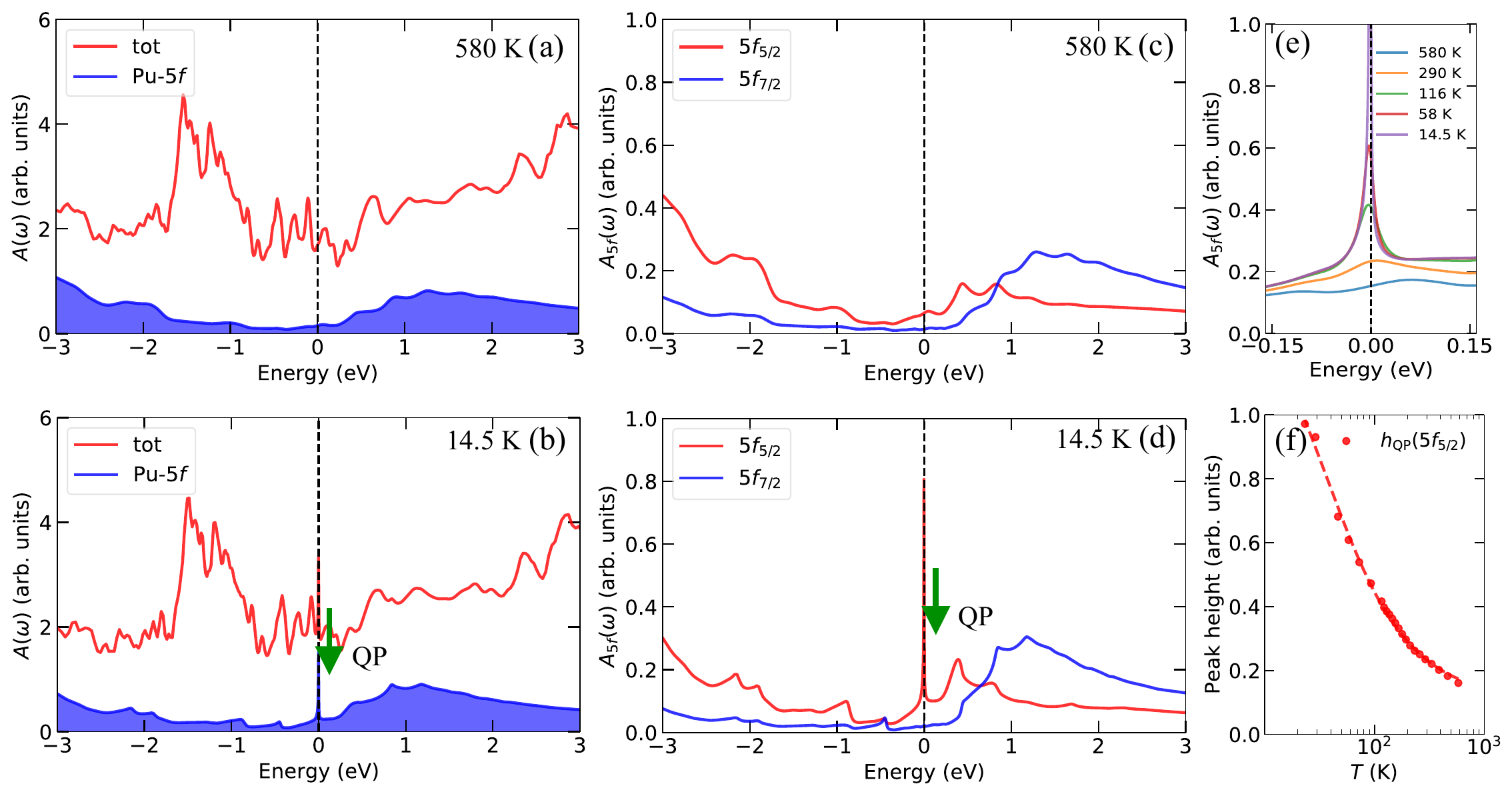}
\caption{(Color online). Electronic density of states of PuIn$_3$.
Total density of states (thick solid lines) and partial 5$f$ density of states (color-filled regions) of 580 K (a), 14.5 K (b). The $j$-resolved 5$f$ partial density of states with $5f_{5/2}$ and $5f_{7/2}$ components represented by red and blue lines, respectively. 580 K (c), 14.5 K (d). (e) The evolution of 5$f$ density of states against temperature in the vicinity of Fermi level. (f) The height of the central quasiparticle peak h$_{\rm QP}$(5$f_{5/2}$) as a function of temperature.
\label{fig:dos}}
\end{figure*}

\begin{figure*}[th]
\centering
\includegraphics[width=\textwidth]{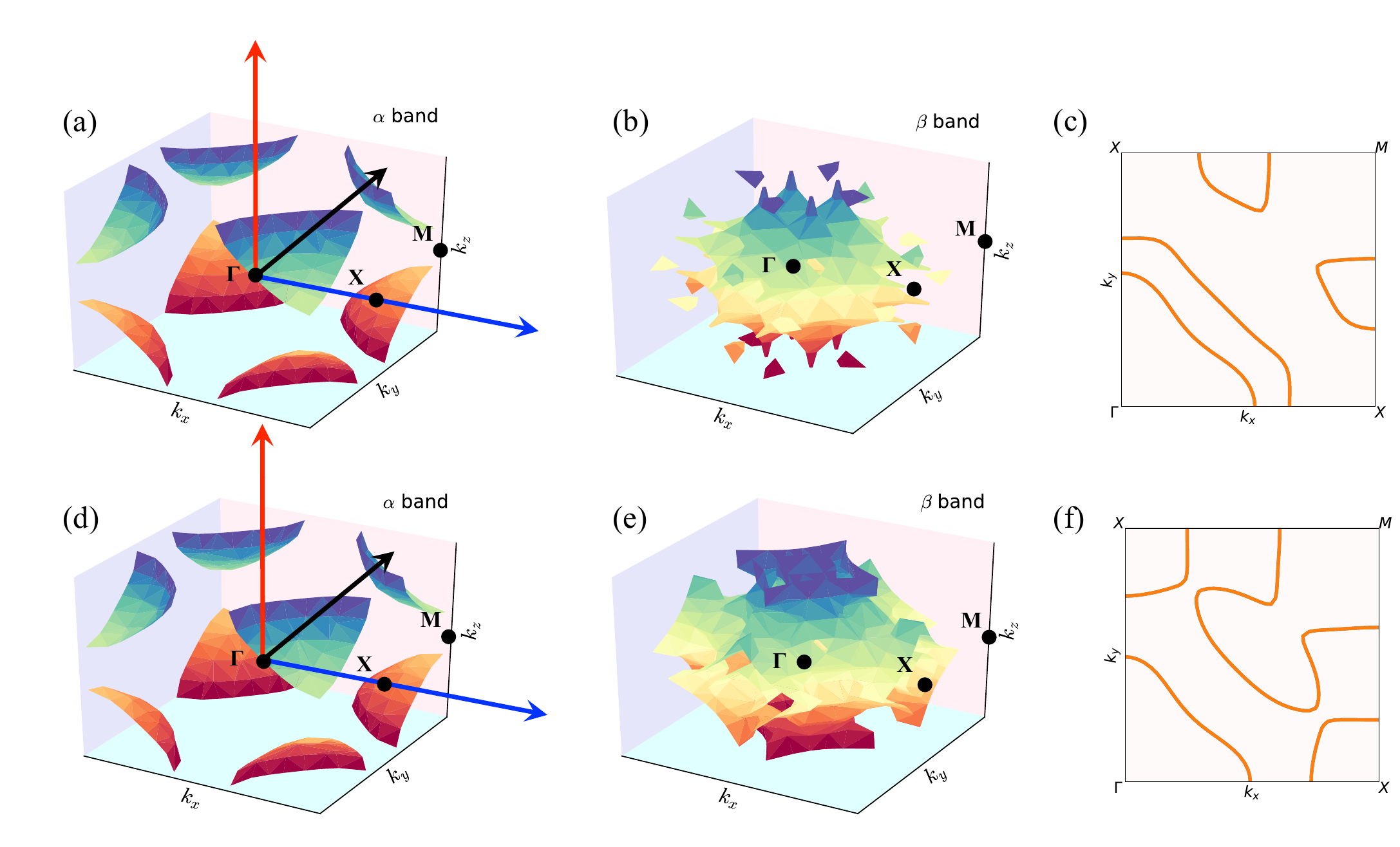}
\caption{(Color online). Three-dimensional Fermi surface and two-dimensional Fermi surface of PuIn$_3$ calculated by the DFT + DMFT method at 580 K (a, b, c) and 14.5 K (d, e, f).
There are two doubly degenerated bands (labelled by $\alpha$ and $\beta$) crossing the Fermi level. 
Three-dimensional Fermi surface of $\alpha$ and $\beta$ bands are plotted in the left and middle columns, respectively. The right columns denote the two-dimensional Fermi surface on the $k_x$-$k_y$ plane ($k_z = \pi/2$) corresponding to $\beta$ bands.
\label{fig:FS}}
\end{figure*}

\begin{figure}[th]
\centering
\includegraphics[width=\columnwidth]{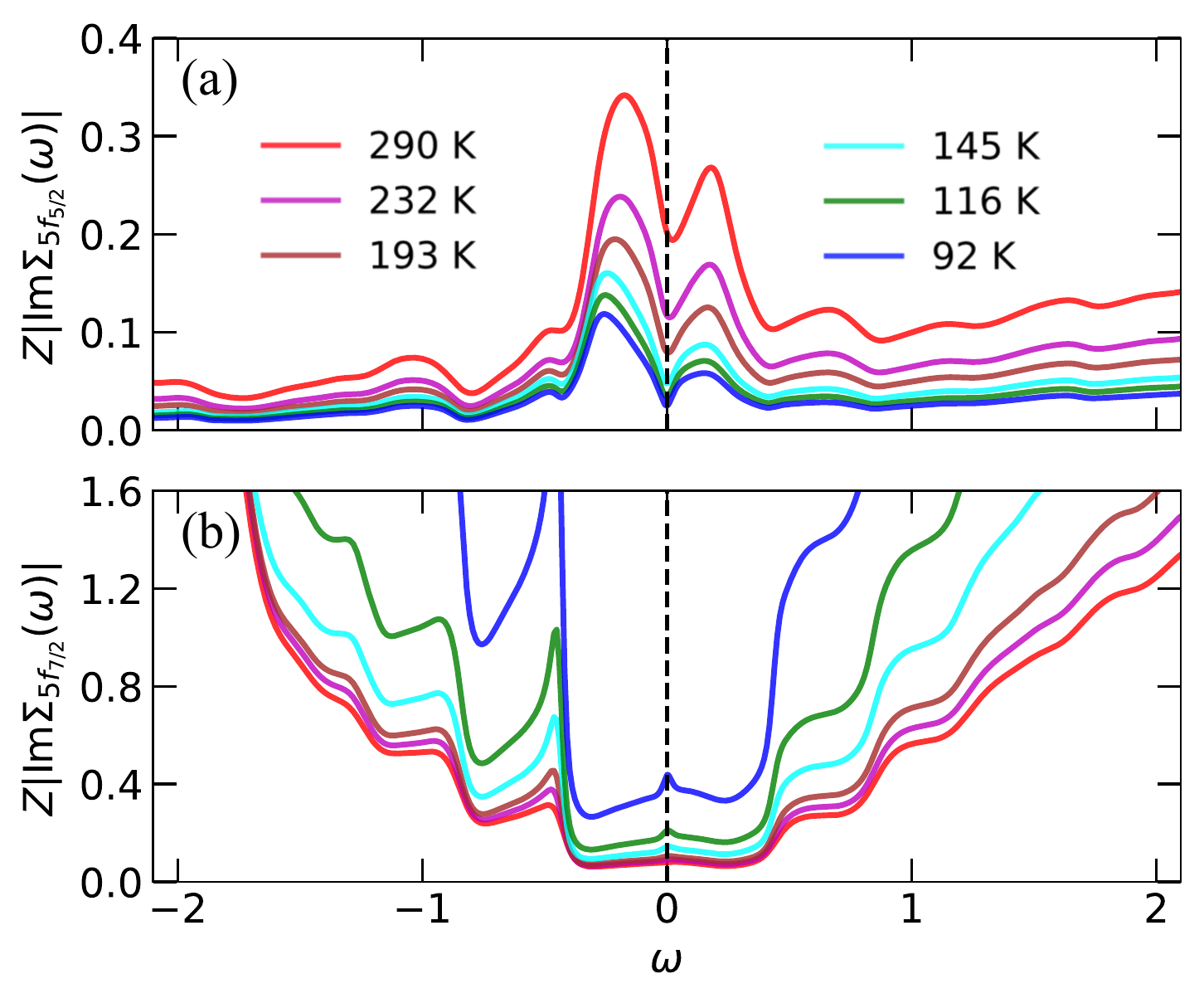}
\caption{(Color online). 
Renormalized real-frequency self-energy functions of PuIn$_3$ obtained by the DFT + DMFT method. (a) and (b) denote temperature-dependent $Z|{\rm Im}\Sigma(\omega)|$ for the $5f_{5/2}$ and $5f_{7/2}$ states, where $Z$ means the renormalization factor. \label{fig:tsigma}} 
\end{figure}

\begin{figure*}[th]
\centering
\includegraphics[width=\textwidth]{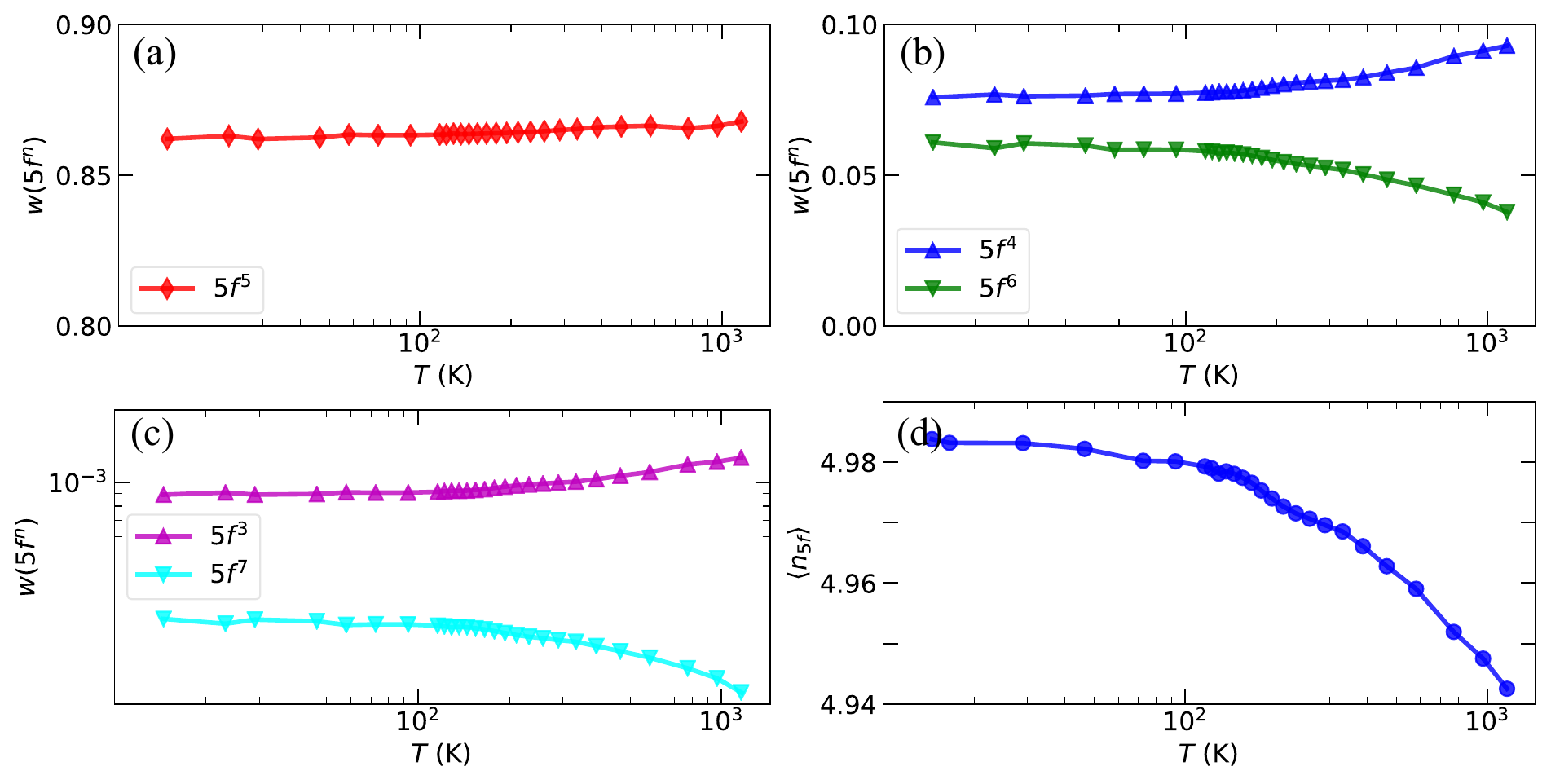}
\caption{(Color online). Probabilities of $5f^{5}$ (red) (a), $5f^{4}$ (blue), $5f^{6}$ (green) configurations (b) and $5f^{3}$ (purple), $5f^{7}$ (cyan) (c), 5$f$ occupancy as a function of temperature (d) for PuIn$_3$ by DFT + DMFT calculations. \label{fig:prob}}
\end{figure*}

\begin{figure*}[th]
\centering
\includegraphics[width=\textwidth]{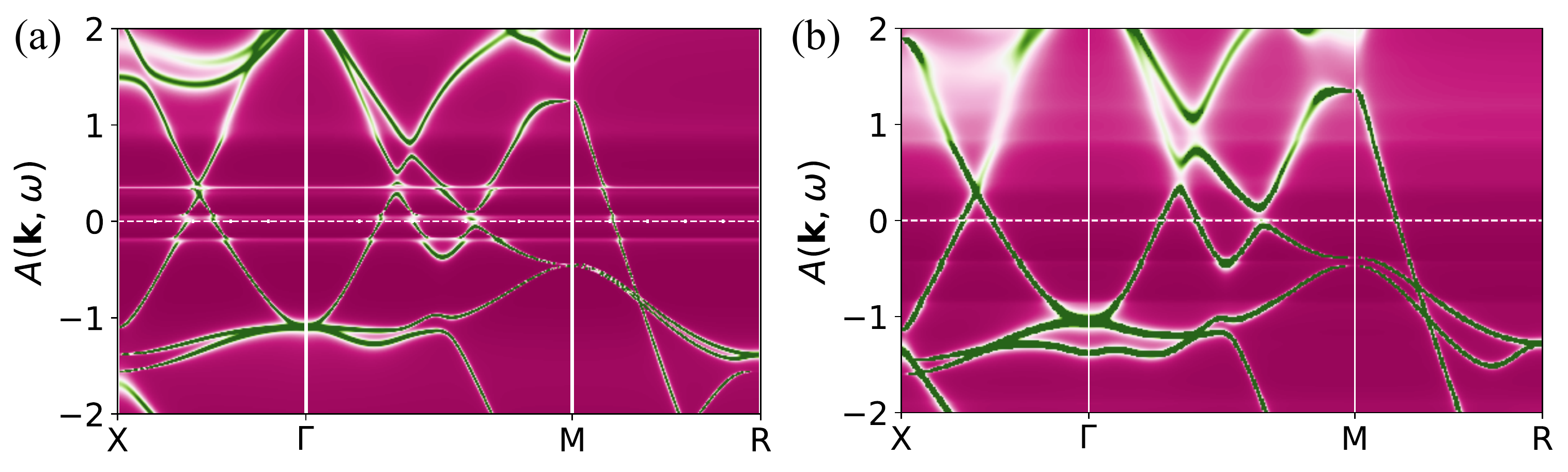}
\caption{(Color online). The momentum-resolved spectral functions $A(\mathbf{k},\omega)$ of both CeIn$_3$ (a) and PuIn$_3$ (b) along the high-symmetry lines in the Brillouin zone obtained by DFT + DMFT method at 14.5 K. The horizontal dashed lines mean the Fermi level. \label{fig:bandcein3}}
\end{figure*}

\subsection{Momentum-resolved spectral functions}
To ascertain the reliability of our calculations, we evaluate the momentum-resolved spectral functions $A(\mathbf{k},\omega)$ of PuIn$_3$ along the high-symmetry lines $X - \Gamma - M - R $ in the irreducible Brillouin zone [see Fig.~\ref{fig:tstruct}(b)] from 580 K to 14.5 K. Figure~\ref{fig:akw} visualizes the temperature dependence of $A(\mathbf{k},\omega)$ at two typical temperatures 580 K and 14.5 K, respectively. In comparison with the available theoretical results by C.-C. Joseph Wang {\it et al.} utilizing a generalized gradient approximation~\cite{PhysRevB.88.125106}, the basic feature of band structures [see Fig.~\ref{fig:akw}(c)] is roughly consistent with each other. For instance, hole-like orbit around -1 eV at $\Gamma$ point and similar hole-like orbit at $M$ point are generally identical. However, the most striking discrepancy is the narrow quasiparticle 5$f$ bands in the vicinity of the Fermi level. It should be pointed out that the enhanced spectral weight of 5$f$ states in the Fermi level at low temperature commonly exists in actinide systems~\cite{PhysRevB.101.125123}. It is speculated that the missing of flat 5$f$ bands in literature~\cite{PhysRevB.88.125106} is partly ascribed to the underestimation of strong correlation among the 5$f$ electrons without fully taking into account the many-body effects.   

After inspecting the detailed characteristics of $A(\mathbf{k},\omega)$ shown in Fig.~\ref{fig:akw}(a) and Fig.~\ref{fig:akw}(c), it is identified that only the conduction bands intersect the Fermi level, indicating the mostly localized 5$f$ states and incoherent quasiparticle bands at high temperature. With gradual decreasing temperature, the overall band profiles generally remain unchanged except for the emergence of flat 5$f$ bands around the Fermi level. In the energy range of -1 eV $\sim$ 1 eV, especially at -0.9 eV and -0.45 eV, there exist nearly dispersionless quasiparticle bands which are associated with 5$f$ states and are split by spin-orbital coupling. At low temperature, 5$f$ states tend to be itinerant and form coherent bands. Along the $X$ - $\Gamma$ and $\Gamma$ - M high-symmetry lines, moderate hybridization between 5$f$ bands and 5$p$ valence electrons are observed. It is guessed that the stripe-like patterns, $c$-$f$ hybridization and coherent quasiparticle bands possibly reveal a temperature-induced localized to itinerant crossover for 5$f$ states in PuIn$_3$.

\subsection{Density of states}
To further explore the electronic structure of PuIn$_3$, we discuss the density of states as a function of temperature in detail. In Fig.~\ref{fig:tdos_exp}, we plot the calculated total density of states and experimental photoemission spectroscopy~\cite{JOYCE2006920} together. Evidently, the representative peaks at the Fermi level and -1.2 eV are correctly confirmed by our results, which serve as critical validation for our calculations. Then Fig.~\ref{fig:dos} shows the electronic density of states at typical temperatures 580 K and 14.5 K. Several features are as follows. 
First of all, the spectral weight at the Fermi level is low, which is almost invisible at high temperature. 
Meanwhile, 5$f$ states are nearly localized to form incoherent states. 
Besides, the broad ``hump" resides from $0.5$ eV to $3$ eV is mainly assigned to the upper Hubbard bands of 5$f$ orbitals and the lower Hubbard bands locate in the energy range of $-3$ eV and $-2$ eV.
Secondly, the spectral weights of upper Hubbard bands and lower Hubbard bands transfer to the Fermi level as temperature lowers. At low temperature, quasiparticle multiplets appear near the Fermi level, manifesting themselves in a pronounced central quasiparticle peak (QP) and two satellite peaks, which are similar to the so called ``Racah materials" like $\delta$-Pu and PuB$_6$. 
Thirdly, owing to the spin-orbital coupling~\cite{RevModPhys.81.235}, the 5$f$ orbitals are split into six-fold degenerated 5$f_{5/2}$ and eight-fold degenerated 5$f_{7/2}$ states. In Fig.~\ref{fig:dos}(d), it is clear that the central quasiparticle peak is mainly constituted by the 5$f_{5/2}$ state and a small satellite peak at -0.45 eV belongs to 5$f_{7/2}$ state, resulting in the energy gap about 0.45 eV. Furthermore, the peak at -1.2 eV is ascribed to 5$f_{5/2}$ state which accords with experimental photoemission spectroscopy. It is worth mentioning that the 5$f_{7/2}$ state remains insulating-like and manifests a gap in the Fermi level. Therefore, the distinguished coherent behavior of 5$f_{5/2}$ state and 5$f_{7/2}$ state is orbital selective.
Fourthly, as is shown in Fig.~\ref{fig:dos}(e), the central quasiparticle peak from 5$f_{5/2}$ states becomes sharp and intense at low temperature. The increment of spectral weight of central quasiparticle peak with decreasing temperature implies the onset of coherent 5$f$ states and appearance of itinerant 5$f$ valence electrons. In consequence, it is roughly concluded that a localized to itinerant crossover may occur with a decline of temperature.

\subsection{Fermi surface topology}
In this subsection, we examine the Fermi surface topology to unveil the temperature-dependent 5$f$ correlated electronic states of PuIn$_3$. The three-dimensional Fermi surface and corresponding two-dimensional Fermi surface at two characteristic temperatures 580 K and 14.5 K are visualized in Fig.~\ref{fig:FS}. It is observed that two doubly degenerated bands intersect the Fermi level (No. of bands: 18 and 19, 16 and 17), which are marked by $\alpha$ and $\beta$, respectively. $\alpha$ bands resemble distorted spherical Fermi surfaces at each of the eight apex angles of the first Brillouin zone, while $\beta$ bands locate at $\Gamma$ point to form ellipsoid shape. As can be seen, the Fermi surface topology of $\alpha$ and $\beta$ bands agree quite well with previous theoretical results~\cite{PhysRevB.88.125106}. Particularly, $\beta$ band corroborates the Fermi surface measured by dHvA experiment~\cite{JPSJ.74.2889}, demonstrating the accuracy of our calculations again. When the temperature goes down, the topology of $\alpha$ bands nearly remains unchanged and the volumes rarely alter either. The key factors lie in $\beta$ bands, which experience topology variation and volume expansion. At high temperature, they cross the $\Gamma$ - $X$ line to formulate ellipsoid-like Fermi surfaces [see Fig.~\ref{fig:FS}(c)] and they intersect the $M$ - $X$ line at low temperature. So the Fermi surface topologies indeed change tremendously with decreasing temperature, which hints the possible Lifshitz transition for 5$f$ states and potential low-temperature magnetic order. The transformation of Fermi surface topology is intimately connected with the temperature-driven localized to itinerant crossover of 5$f$ correlated electronic states, which could be detected using quantum oscillation measurements~\cite{PhysRevB.88.125106}.

\subsection{Self-energy functions}
As mentioned above, the 5$f$ electrons are strongly correlated and the electron correlation effects can be deduced from their electron self-energy functions~\cite{RevModPhys.68.13,RevModPhys.78.865}. Figure~\ref{fig:tsigma} illustrates the renormalized imarginary part of self-energy functions $Z|{\rm Im}\Sigma(\omega)|$ for 5$f_{5/2}$ and 5$f_{7/2}$ states.  
Here $Z$ means the quasiparticle weight or renormalization factor, which denotes the electron correlation strength and can be obtained from the real part of self-energy functions via the following equation~\cite{RevModPhys.68.13}:
\begin{equation}
Z^{-1} = \frac{m^\star}{m_e} = 1 - \frac{\partial \text{Re} \Sigma(\omega)}{\partial \omega} \Big|_{\omega = 0}. \label{eqsigma}
\end{equation}
Generally, $Z|{\rm Im}\Sigma(0)|$ is considered as an indicator of low-energy electron scattering rate~\cite{PhysRevB.99.125113}. At low temperature, $Z|{\rm Im}\Sigma_{5f_{5/2}}(0)|$ approaches zero, demonstrating the itinerant nature of 5$f$ states. With elevating temperature, $Z|{\rm Im}\Sigma_{5f_{5/2}}(\omega)|$ rises swiftly, especially in the low-energy regime of [-0.5 eV, 0.5 eV], and then reaches finite values. Conversely, $Z|{\rm Im}\Sigma_{5f_{7/2}}(\omega)|$ surges up quickly at high-energy regime ($|\omega|$ $<$ 0.5 eV). So the enhancement of $Z|{\rm Im}\Sigma_{5f_{7/2}}(\omega)|$ becomes more significant than that of $Z|{\rm Im}\Sigma_{5f_{5/2}}(\omega)|$, which leads to the suppression for the itinerancy of 5$f_{7/2}$ state and explains its energy gap in the Fermi level. Since self-energy functions of 5$f_{5/2}$ and 5$f_{7/2}$ states manifest differentiated temperature-dependent patterns, it is concluded that 5$f$ electron correlation are orbital selective.

\subsection{Atomic eigenstate probabilities}
In analogy with the archetypal mixed-valence metal $\delta$-Pu~\cite{shim:2007} whose average 5$f$ electron occupation deviated from its nominal value 5.0, PuIn$_3$ is expected to display mixed-valence behavior since it shares some common features like the three-peak structure in the spectral functions of $\delta$-Pu. To interpret the valence state fluctuations and mixed-valence behavior, we attempt to obtain the 5$f$ electron atomic eigenstates from the output of DMFT ground states. Here $p_\Gamma$ is adopted to quantify the probability of 5$f$ electrons which stay in each atomic eigenstate $\Gamma$. Then the average 5$f$ valence electron is expressed as $\langle n_{5f} \rangle = \sum_\Gamma p_\Gamma n_\Gamma$, where $n_\Gamma$ is the number of electrons in each atomic eigenstate $\Gamma$. Finally, the probability of 5$f^n$ electronic configuration can be defined as $\langle w(5f^{n}) \rangle = \sum_\Gamma p_\Gamma \delta (n-n_\Gamma)$. 

Figures~\ref{fig:prob}(a)-(c) depict the calculated probability of 5$f^n$ electronic configuration for PuIn$_3$, where $n \in$ [3, 7] and other probability of electronic configurations are too small to be seen. As listed in Table~\ref{tab:prob}, the probability of 5$f^5$ electronic configuration is overwhelmingly dominant, which accounts for 86\%, followed by 5$f^4$ and 5$f^6$ electronic configurations. At 580 K, the probability of 5$f^4$ and 5$f^6$ electronic configurations stand at 8.6\% and 4.7\%, respectively. In the meantime, the occupation of 5$f$ electrons is approximately 4.96, which approaches its nominal value 5.0. Hence 5$f$ electrons are inclined to stay more time at 5$f^5$ electronic configuration than 5$f^4$ and 5$f^6$ electronic configurations. It means that valence state fluctuations are not quite remarkable at relatively high temperature, signifying localized 5$f$ states. As the temperature lowers, the probability of 5$f^5$ electronic configuration slightly decreases, accompanied by a subtle deduction of 5$f^4$ and a minor growth of 5$f^6$ electronic configuration. Overall, the probability of 5$f^n$ electronic configurations are not sensitive to varying temperature. As a consequence, the occupation of 5$f$ electrons keeps almost unchanged, suggesting the atypical mixed-valence behavior of PuIn$_3$, which is beyond our expectation. In this respect, the temperature-induced localized to itinerant crossover of 5$f$ states is hidden in the atomic eigenstate probabilities.

\begin{table}[th]
\caption{Probabilities of $5f^{3}$, $5f^{4}$, $5f^{5}$, $5f^{6}$, and $5f^{7}$ for PuIn$_3$ at temperatures 580 K and 14.5 K, respectively.
 \label{tab:prob}}
\begin{ruledtabular}
\begin{tabular}{cccccc}
Temperatures & $5f^{3}$ & $5f^{4}$ & $5f^{5}$ & $5f^{6}$ & $5f^{7}$ \\
\hline
580 K    & 1.106$\times 10^{-3}$ & 0.086  & 0.866 & 0.047 & 1.894$\times 10^{-4}$ \\
14.5 K   & 8.917$\times 10^{-4}$ & 0.076  & 0.862 & 0.061 & 2.738$\times 10^{-4}$ \\
\end{tabular}
\end{ruledtabular}
\end{table}

\section{Discussion\label{sec:dis}}
\subsection{$f$ electrons in CeIn$_3$ and PuIn$_3$}

Here we concentrate on the isostructural compounds CeIn$_3$ and PuIn$_3$ to unravel the alluring $f$ electrons and fascinating bonding behavior. As mentioned above, CeIn$_3$ and PuIn$_3$ stabilize in cubic AuCu$_3$ structure with similar lattice constant 4.689 \AA~\cite{BENOIT1980293} and 4.703 \AA~\cite{Eric2466}, respectively. 
Since the element identification is regarded as the chemical substitution Ce for Pu, the electronic structure and related physical properties are expected to share abundant common traits. For instance, CeIn$_3$ and PuIn$_3$ are typical heavy-fermion compounds~\cite{PhysRevB.65.024425,Bauer2015Plutonium}, which develop antiferromagnetic order below Ne\'{e}l temperature 10 K~\cite{PhysRevB.22.4379} and 14.5 K~\cite{Chudo2013}, respectively.

Figure~\ref{fig:bandcein3} presents the calculated momentum-resolved spectral functions $A(\mathbf{k},\omega)$ of both CeIn$_3$ and PuIn$_3$ along the same high-symmetry lines in the Brillouin zone via DFT + DMFT method at 14.5 K. Several features are as follows. Firstly, the parallel flat bands at 0.1 eV and 0.4 eV [see Fig.~\ref{fig:bandcein3}(a)] are attributed to Ce-4$f$ electrons, which are split by spin-orbital coupling into $j = 5/2$ and $j = 7/2$ subbands with energy separation being approximately 0.3 eV. Secondly, the narrow 4$f$ bands intersect conduction bands to form $c-f$ hybridization and open obvious hybridization gaps. Thirdly, the electron-like band and hole-like band only adjoin along the $X$ - $\Gamma$ line in the angle-resolved photoemission spectroscopy experiment~\cite{yun:2015}, which is well reproduced by our calculation. Fourthly, the conduction bands with strong energy dispersions are mainly contributed by In atoms. Consequently, the electron-like band and hole-like band at $\Gamma$ and $M$ points located about -1 eV and -0.5 eV below the Fermi level, respectively. 

For comparison, it is significative to evaluate band structure of PuIn$_3$ [see Fig.~\ref{fig:bandcein3}(b)].
First of all, the overall energy profiles seem incredibly similar for CeIn$_3$ and PuIn$_3$, even though the band degeneracy at $\Gamma$ point in CeIn$_3$ is lifted in PuIn$_3$. Such amazing similarity in the band structure is attributed to the conduction band of In atoms. Secondly, the dispersionless flat 5$f$ bands mainly distribute at -0.9 eV and -0.45 eV below the Fermi level, where the spin-orbital coupling energy separation between $j = 5/2$ and $j = 7/2$ states of 5$f$ electrons is about 0.45 eV. It is reasonable that the stronger spin-orbital coupling strength of Pu results in a larger energy separation than that of Ce, since the atomic number of Pu is much higher than that of Ce. Thirdly, both 4$f$ and 5$f$ states are strongly correlated with strikingly renormalized bands and electron effective masses. Lastly, it is widely believed that 4$f$ states are more localized than the 5$f$ states.
However, 4$f$ states in CeIn$_3$ seem to take in active chemical bonding and becomes itinerant at low temperature. Conversely, the 5$f$ states in PuIn$_3$ undergoes a hidden localized-itinerant crossover at low temperature. In this scenario, the $f$ electron nature in CeIn$_3$ and PuIn$_3$ pose archetypical prototype to gain deep understanding on the intrinsic connection between 4$f$ and 5$f$ states, so as to uncover low temperature antiferromagnetic order.

\subsection{dHvA quantum oscillation}
dHvA quantum oscillation is known as a useful tool to detect the Fermi surface.  
The dHvA effect encodes the magnetic field dependence of the quantum oscillations in magnetization and other properties owing to the change in the occupation of Landau levels driven by the oscillation of the magnetic field. An oscillation frequency proportional to the extremal Fermi surface cross-sectional area perpendicular to the magnetic field direction is expressed as:
\begin{equation}
F = \frac{\hbar}{2\pi e} A, \label{dHvAfreq}
\end{equation}
where $F$ is the dHvA frequency in unit of kT, $e$ is the elementary charge, $\hbar$ is the reduced Planck constant, $A$ denotes the extremal area. Moreover, the electron effective mass averaged around the extremal orbits can be obtained from the damping strength as a function of temperature. Thus the dHvA frequency and electron effective mass are evaluated using the numerical algorithm implemented by Rourke and Julian~\cite{ROURKE2012324}. In comparison with dHvA experimental results, the magnetic field is chosen along the [111] direction. The calculated dHvA frequency together with electron effective mass are listed in Table~\ref{tab:dHvAfreq}. 
The calculated dHvA frequency of $\beta$ band is 1.91 kT, which is close to the experimental value about 2.0 kT and theoretical value 2.18 kT~\cite{JPSJ.74.2889} based on 5$f$-itinerant band model. However, our calculated dHvA frequency differs from the value using a generalized gradient approximation~\cite{PhysRevB.88.125106}. These discrepancies might arise from the approximations of DFT calculation methods.
Furthermore, the obtained electron effective mass is 1.45 $m_e$, approaching the previous theoretical value 1.56 $m_e$~\cite{JPSJ.74.2889}. 
In view of the accordance between the theoretical and experimental results, it is guessed that the electronic correlations indeed affects the Fermi surface and electron effective mass, because the discrepancy exists between DFT results and experimental dHvA frequencies for paramagnetic state of PuIn$_3$. Additionally, the other orbits are not discussed here, which might serve as critical prediction for further dHvA experiment.

\begin{table}[th]
\caption{The calculated dHvA frequency and electron effective mass of PuIn$_3$ with magnetic field along the [111] direction. h means hole-like orbit, e denotes electron-like orbit.
\label{tab:dHvAfreq}}
\begin{ruledtabular}
\begin{tabular}{ccc}
F (kT) & $m^* (m_e)$ & Type \\
\hline
1.91   & 1.45  & e \\
5.33   & 4.62  & e \\
7.68   & 6.57  & e \\
9.10   & 7.39  & h \\
\end{tabular}
\end{ruledtabular}
\end{table}


\section{conclusion\label{sec:summary}}
In summary, we studied the detail electronic structures of PuIn$_3$ by employing a state-of-the-art first-principles many-body approach. The temperature dependence of itinerant to localized crossover and the correlated electronic states were addressed systematically. As the temperature declines, the augmented itinerancy of 5$f$ electrons and emergence of valence state fluctuations indicate a localized-itinerant crossover. Especially, 5$f$ states manifest orbital selective electron correlation, reflected by orbital-dependent electron effective masses and renormalized bands.
Our results not only provide a comprehensive picture about the evolution of 5$f$ correlated electronic states with respect to temperature, but also gain important implications into the low temperature magnetism in PuIn$_3$. Further studies about the other Pu-based compounds are undertaken.

\begin{acknowledgments}
This work was supported by the National Natural Science Foundation of China (No.~11704347, No.~11874329, No.~11934020, No.~22025602), and the Science Challenge Project of China (No.~TZ2016004). 
\end{acknowledgments}


\bibliography{PuIn3}

\end{document}